\documentclass[
aps, 
prl, 
onecolumn,
twocolumn,
10pt,
superscriptaddress
]{revtex4-2}

\usepackage{amsmath}
\usepackage{graphicx}
\usepackage{xcolor}
\usepackage{setspace}
\usepackage{physics}

\usepackage{etoolbox}
\makeatletter
	\newcommand{\appendixformat}{
		\setcounter{secnumdepth}{2}
		\setcounter{section}{0}
		\setcounter{figure}{0}
		\setcounter{table}{0}
		\renewcommand{\thefigure}{S\arabic{figure}}
		\renewcommand{\thetable}{S\arabic{table}}
	}

\begin{document}
		
		
		\title{Nuclear quantum effects induce superionic proton transport in nanoconfined water}
		
		\author{Pavan Ravindra}
		\affiliation{%
			Department of Chemistry, Columbia University, 3000 Broadway, New York, NY 10027, USA
		}
		\affiliation{%
			Yusuf Hamied Department of Chemistry, University of Cambridge, Lensfield Road, Cambridge, CB2 1EW, UK
		}
		
		\author{Xavier R. Advincula}
		\affiliation{%
			Yusuf Hamied Department of Chemistry, University of Cambridge, Lensfield Road, Cambridge, CB2 1EW, UK
		}
		\affiliation{%
			Cavendish Laboratory, Department of Physics, University of Cambridge, Cambridge, CB3 0HE, UK
		}
		\affiliation{%
			Lennard-Jones Centre, University of Cambridge, Trinity Ln, Cambridge, CB2 1TN, UK
		}

		\author{Benjamin X. Shi}
		\affiliation{%
			Yusuf Hamied Department of Chemistry, University of Cambridge, Lensfield Road, Cambridge, CB2 1EW, UK
		}
		\affiliation{%
			Lennard-Jones Centre, University of Cambridge, Trinity Ln, Cambridge, CB2 1TN, UK
		}
		
		\author{Samuel W. Coles}
		\affiliation{%
			Yusuf Hamied Department of Chemistry, University of Cambridge, Lensfield Road, Cambridge, CB2 1EW, UK
		}
		\affiliation{%
			Lennard-Jones Centre, University of Cambridge, Trinity Ln, Cambridge, CB2 1TN, UK
		}
		
		\author{Angelos Michaelides}%
		\email{am452@cam.ac.uk}
		\affiliation{%
			Yusuf Hamied Department of Chemistry, University of Cambridge, Lensfield Road, Cambridge, CB2 1EW, UK
		}
		\affiliation{%
			Lennard-Jones Centre, University of Cambridge, Trinity Ln, Cambridge, CB2 1TN, UK
		}
		
		\author{Venkat Kapil}%
		\email{v.kapil@ucl.ac.uk}
		\affiliation{%
			Yusuf Hamied Department of Chemistry, University of Cambridge, Lensfield Road, Cambridge, CB2 1EW, UK
		}
		\affiliation{%
			Lennard-Jones Centre, University of Cambridge, Trinity Ln, Cambridge, CB2 1TN, UK
		}
		\affiliation{%
			Department of Physics and Astronomy, University College London, 7-19 Gordon St, London WC1H 0AH, UK
		}
		\affiliation{%
			Thomas Young Centre and London Centre for Nanotechnology, 9 Gordon St, London WC1H 0AH
		}

		\begin{abstract}
			Recent work has suggested that nanoconfined water may exhibit superionic proton transport at lower temperatures and pressures than bulk water.
			Using first-principles-level simulations, we study the role of nuclear quantum effects in inducing this superionicity in nanoconfined water.
			We show that nuclear quantum effects increase the ionic conductivity of nanoconfined hexatic water, leading to superionic behaviour at lower temperatures and pressures than previously thought possible.
			Our work suggests that superionic water may be accessible in graphene nanocapillary experiments.
		\end{abstract}
		
		
		\maketitle
		
		\newpage
		
		
		%
		Despite its simple molecular structure, water exhibits complex phase behaviours. The known phases of bulk water include over fifteen crystalline ice phases~\cite{salzmann_advances_2019}, several amorphous ice phases~\cite{burton_crystal_1997, loerting_second_2001, nelmes_annealed_2006, rosu-finsen_medium-density_2023}, possibly two distinct liquid phases in the supercooled regime~\cite{kim_experimental_2020}, and superionic phases characterized by protons diffusing through a fixed oxygen sub-lattice~\cite{husband2024phase,millot_experimental_2018, millot_nanosecond_2019,sun_phase_2015}.
		Recent studies have brought particular attention to the properties of water when it is confined to nanometer-scale cavities, as its behaviour changes dramatically from the bulk in these conditions~\cite{secchi_massive_2016, gopinadhan_complete_2019, fumagalli_anomalously_2018,kavokine_fluctuation-induced_2022}.
		Perhaps unsurprisingly, the phase diagram of water remains remarkably complex in nanoconfined environments~\cite{algara-siller_square_2015,kapil_first-principles_2022, lin_temperature-pressure_2023,jiang_rich_2024,li_replica_2019,giovambattista_computational_2012}.
		The predicted stable phases are generally made up of 1-3 layers of water molecules with microscopic structures substantially different from bulk ice phases.~\cite{corsetti_structural_2016,ravindra_quasi-one-dimensional_2023}
		Furthermore, nanoscale confinement can induce phases that qualitatively differ from the bulk, such as the hexatic phase—an intermediate state between solid and liquid, as predicted by the Kosterlitz-Thouless-Halperin-Nelson-Young (KTHNY) theory~\cite{kapil_first-principles_2022, zubeltzu_continuous_2016}.
		Traditional empirical force fields have provided a valuable starting point towards understanding water in confined systems~\cite{algara-siller_square_2015,li_replica_2019,han_phase_2010,johnston_liquid_2010,kumar_thermodynamics_2011,zubeltzu_continuous_2016,corsetti_structural_2016}.
		However, since these force fields are parameterized to reproduce bulk properties specifically, it remains unclear if they are consistently reliable for confined water~\cite{li_replica_2019,chen_evidence_2016}.
		Accordingly, recent studies have instead employed first-principles calculations to explore the stable phases of nanoconfined water~\cite{zubeltzu_continuous_2016,corsetti_structural_2016,chen_two_2016}.
		Machine learning interatomic potentials (MLIPs) have allowed for the accuracy of these first-principles studies to be extended to finite temperature conditions~\cite{behler_constructing_2015,schran_committee_2020}.
		MLIPs have been used to predict phase diagrams of mono-, bi-, and tri-layer films of water in graphene-like cavities~\cite{kapil_first-principles_2022, lin_temperature-pressure_2023, ravindra_quasi-one-dimensional_2023, jiang2021first, jiang_rich_2024}.
		These studies generally consider confinement widths of 5-8\,\AA{}, and they have shown that the phase behaviour of nanoconfined water changes substantially over this interval~\cite{kapil_first-principles_2022,lin_temperature-pressure_2023,jiang2021first}.
		An intriguing outcome of nano-scale confinement is the reduction in the onset conditions for superionic behaviour in water~\cite{kapil_first-principles_2022,jiang_rich_2024}.
		In superionic phases of water, oxygen atoms remain in their lattice positions, while protons are able to hop from oxygen to oxygen.
		This leads to unusually large ionic conductivities as a result of liquid-like proton diffusion.
		Bulk water has been known to exhibit superionic proton transport at extreme temperature and pressure conditions on the order of hundreds of GPa and thousands of Kelvin~\cite{millot_experimental_2018, millot_nanosecond_2019}.
		As these conditions are only relevant in extreme settings, such as in the interiors of icy giant planets, superionic proton transport has remained inaccessible to practical conditions.
		However, \citet{kapil_first-principles_2022} have predicted that monolayer water in a 5\,\AA{} wide graphene-like cavity can exhibit superionic behaviour on the order of a few GPa and a few hundred Kelvin above room temperature.
		~\citet{jiang_rich_2024} extended this study to the 6-8\,\AA{} range and identified new superionic nanoconfined water phases.
		This work also stressed the importance of including nuclear quantum effects (NQEs) when considering proton dynamics.
		The superionic phases in Ref.~\citenum{jiang_rich_2024} were observed at dramatically reduced temperatures and pressures, but the onset conditions were still on the order of tens of GPa and 600-800\,K.
		Although still relevant in extreme environments, these temperatures and pressures fall outside of the range of conditions that are accessible to modern nanoconfined water experiments~\cite{algara-siller_square_2015}.

		In this Letter, we demonstrate that NQEs significantly reduce the onset conditions for superionic proton transport in monolayer water for a 5\,\AA{} confinement width.
		NQEs enhance O--H dissociation in both the solid and hexatic phases of nanoconfined water to a degree beyond what is observed in bulk or even at wider confinement widths.
		In the hexatic phase, we observe superionic behaviour at lateral pressures on the order of 1\,GPa.
		Importantly, this is the same pressure range as in graphene nanocapillary experiments~\cite{algara-siller_square_2015}.
		The large lateral pressure in these experiments arises due to the van der Waals attraction that pulls the graphene sheets together~\cite{algara-siller_square_2015, jiao_structures_2017}.
		This brings forward the possibility for detecting superionic water in lab conditions and potentially realizing a new class of nano-devices that leverage superionicity to achieve novel behaviours.
		%
		
		
		%
		\textit{Methods.} Our simulation setup mimics the conditions experienced by a monolayer film of water trapped between two sheets of graphene~\cite{algara-siller_square_2015}.
		We model the water-carbon interactions using a uniform, atomistically flat confining potential that is fit to calculations performed at the quantum Monte Carlo (QMC) level~\cite{chen_two_2016}.
		Such a setup has been used in previous simulation studies of graphene-like nanocapillaries~\cite{chen_two_2016, li_replica_2019, kapil_first-principles_2022, lin_temperature-pressure_2023, jiang_rich_2024} and has also been compared to setups containing explicit carbon atoms~\cite{jiao_structures_2017,algara-siller_square_2015,corsetti_structural_2016}.
		We incorporate water-water interactions at the revPBE0-D3 level using the Behler-Parrinello MLIP~\cite{behler_constructing_2015} from Ref.~\citenum{kapil_first-principles_2022}.
		revPBE0-D3 is a dispersion-corrected hybrid density functional that has previously been validated against QMC calculations of confined ice~\cite{chen_evidence_2016}.
		In Section~\ref{sisec:benchmark} of the Supplementary Materials, we further assess the accuracy of the revPBE0-D3 functional for reactive events.
		We do this by estimating the potential energy barrier for proton transfer in monolayer superionic water at the level of coupled cluster theory with single, double, and perturbative triple excitations (CCSD(T)) with a local approximation.
		We observe a good performance of the revPBE0-D3 functional, as it achieves better than chemical accuracy with respect to CCSD(T) on the per monomer forward reaction barrier.
		Our trained MLIP is able to match the performance of revPBE0-D3 on the forward barrier as well, which ensures that our MLIP faithfully reproduces the results of the underlying functional during O--H dissociation events.
		Classical and path integral MLIP simulations are performed using \texttt{i-PI v2.0}~\cite{kapil_i-pi_2019} and the \texttt{LAMMPS-n2p2} interface~\cite{singraber_library-based_2019, thompson_lammps_2022}.  
		Equilibrium properties, such as free energy profiles, are estimated by averaging over the replicas of the path integral simulations.
		Dynamical properties, such as mean squared displacements and ionic conductivities, are estimated using the dynamics of the centroids~\cite{cao_new_1993}.
		To reduce the high cost of centroid molecular dynamics, we employ the recently developed Path-Integral coarse-Grained Simulations (PIGS) technique~\cite{musil_quantum_2022}.
		This method involves training a secondary MLIP to fit the centroid potential of mean force by learning a correction to the original potential energy surface. 
		By performing standard molecular dynamics on the learned centroid potential of mean force, we are able to run centroid molecular dynamics at the same cost as classical simulations.
		The PIGS technique boosts the computational efficiency of our simulations by over two orders of magnitude, allowing us to efficiently estimate the quantum transport properties reported in this work.
		Further details on the computational settings can be found in Section~\ref{sisec:comp_details} of the Supplementary Materials.

		\begin{figure*}[!t]
			\centering
			\includegraphics[width=\textwidth]{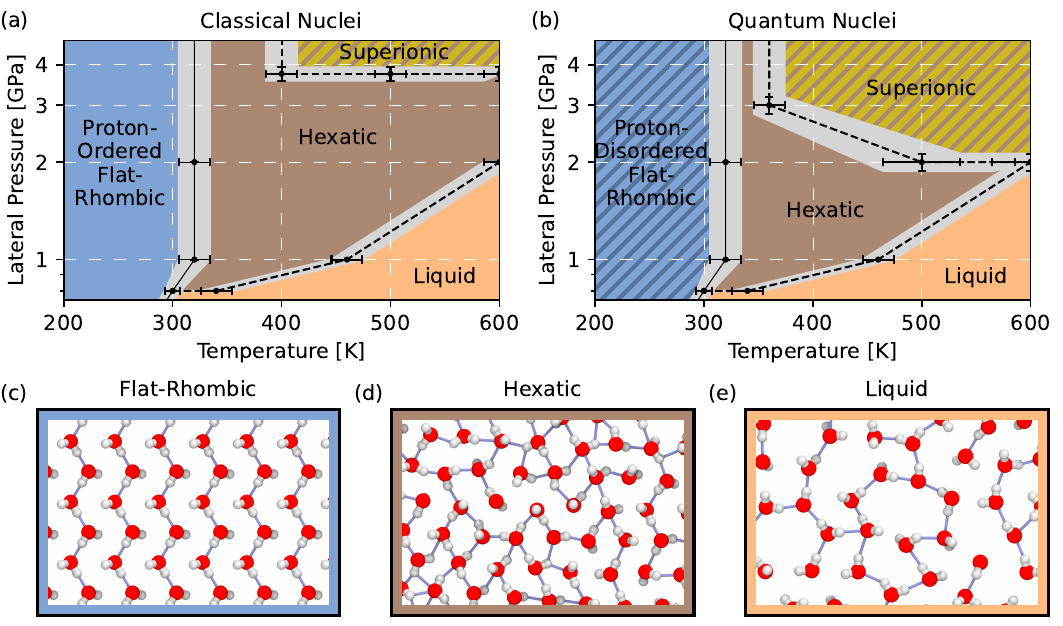}
			\caption{\textbf{Impact of nuclear quantum effects on the phase diagram of nanoconfined water.} (a) A portion of the phase diagram of nanoconfined water, computed with classical nuclei. The grey regions represent the regions of uncertainty associated with the phase boundaries. (b) The same portion of the phase diagram with nuclear quantum effects taken into account. The detailed procedure for computing these phase diagrams is provided in Section~\ref{sisec:comp_details} of the Supplementary Materials. (c), (d), (e) Snapshots of the (proton-ordered) flat-rhombic phase, hexatic phase, and liquid phase, respectively. These snapshots come from our simulations with classical nuclei, and they show only a zoomed-in portion of the simulation cell.}
			\label{fig:phase_diagram}
		\end{figure*}
		
		
		%
		\textit{Results.} In Fig.~\ref{fig:phase_diagram}(a), we report a part of the phase diagram from Ref.~\citenum{kapil_first-principles_2022} spanning the 1-4 GPa lateral pressure range.
		We chose this pressure range because it is on the same order of magnitude as the van der Waals pressure in graphene nanocapillary experiments~\cite{algara-siller_square_2015}.
		In this region of the phase diagram, the stable phase at low temperatures is the flat-rhombic ice phase, depicted in Fig.~\ref{fig:phase_diagram}(c).
		KTHNY theory for two-dimensional phase transitions predicts that two-dimensional solid phases should melt through a two-step melting process~\cite{kosterlitz_long_1972,kosterlitz_ordering_1973}.
		This two-step mechanism can be seen in Fig.~\ref{fig:phase_diagram}(a) along the 1 GPa lateral pressure isobar.
		Along this isobar, the flat-rhombic phase undergoes an initial phase transition to a hexatic phase at around 320\,K, followed by a second phase transition to a liquid phase around 460\,K.
		One hallmark of the intermediate hexatic phase is the six-fold orientational order in its oxygen sub-lattice, which can be seen in Fig.~\ref{fig:phase_diagram}(d).
		This six-fold orientational order is lost in the liquid phase, shown in Fig.~\ref{fig:phase_diagram}(e).
		This phase diagram also shows that the hexatic phase begins to exhibit superionic proton transport at higher temperatures and pressures, indicated by the hatched yellow region.
		In this superionic hexatic phase, the hexatic oxygen sub-lattice remains fixed, but protons begin to readily diffuse through the hexatic sub-lattice. 
		In Fig.~\ref{fig:phase_diagram}(b), we report the changes in phase behaviours when NQEs are included.
		It is well known that NQEs do not activate the long-ranged and low-energy intermolecular modes that induce a global reordering of molecules in water~\cite{markland_refined_2008, habershon_competing_2009,reinhardt_quantum-mechanical_2021}.
		Consistent with this notion and our previous work on nanoconfined water~\cite{kapil_first-principles_2022}, we do not observe changes across the solid-to-liquid phase transitions with the inclusion of NQEs.
		In contrast, we observe large changes in the phase behaviours of the flat-rhombic, hexatic, and superionic water due to increased delocalization of protons.
		In bulk phases, superionic proton transport and proton disorder in high-density ice arise from intramolecular O--H fluctuations and are known to be affected by the zero-point motion of protons~\cite{sun_phase_2015, cheng_phase_2021, benoit_tunnelling_1998}. 
		While these changes to bulk phase behaviours occur at tens or hundreds of GPa, the changes that we observe under nanoconfinement happen at much lower lateral pressures in the 1-4\,GPa range that we focus on in this work.
		%
		
		
		\begin{figure}[t!h]
			\centering
			\includegraphics[width=3.375in]{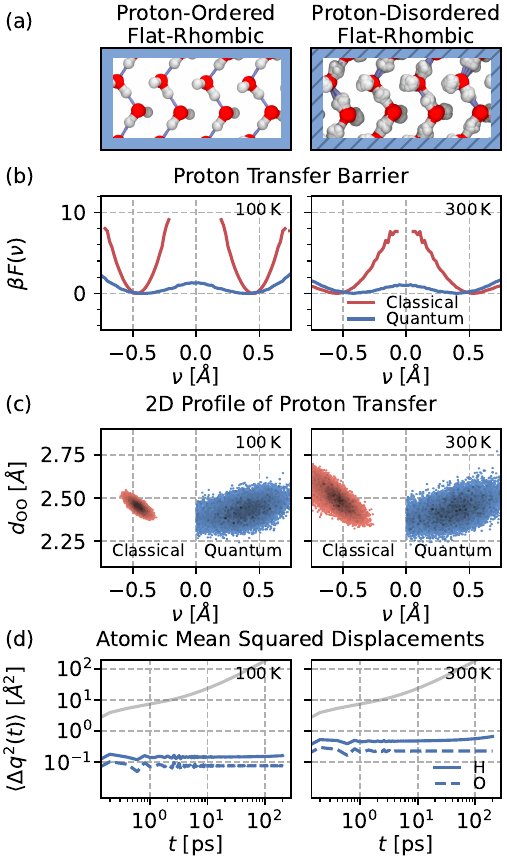}
			\caption{\textbf{Proton disorder in nanoconfined ice.} (a) Snapshots of the flat-rhombic phase without (left) and with (right) nuclear quantum effects. These snapshots illustrate proton order and disorder, respectively. The snapshot on the right shows all 32 beads from our path integral molecular dynamics simulations, overlaid on top of each other. (b) Free energy profiles along the proton transfer coordinate $\nu$ for the flat-rhombic phase without (red) and with (blue) nuclear quantum effects at 100\,K (left) and 300\,K (right). (c) The joint free energy profile of neighbouring oxygen-oxygen distances $d_{\text{OO}}$ and the proton transfer coordinate $\nu$ for the flat-rhombic phase. Darker regions correspond to lower free energies. By definition, the distributions are symmetric about $\nu$, so we only show the classical (red) distributions for negative $\nu$ values and the quantum (blue) distributions for positive $\nu$ values. (d) The mean squared displacement of protons (solid lines) and oxygen atoms (dashed lines). An example of the characteristic diffusive mean squared displacement for the superionic phase is shown in grey for reference. This grey curve is taken from path integral molecular dynamics simulations of the superionic hexatic phase at 600\,K.}
			\label{fig:proton_disorder}
		\end{figure}

		To determine whether NQEs induce proton transfer in nanoconfined water, we compute the free energy profiles along the proton transfer coordinate $\nu$ for the protons in our simulations.
		The proton transfer coordinate $\nu$ is defined such that protons with large absolute values of $\nu$ are much closer to their nearest oxygen atom than they are to their second nearest oxygen atom.
		Conversely, protons with $\nu$ near 0 are equidistant from their two nearest oxygen atoms.
		In order for a proton to transfer from one oxygen atom to another, it must pass through $\nu=0$.
		Hence, the height of the free energy barrier at $\nu=0$ provides information about the rate of proton transfer between oxygen atoms.
		The mathematical definition of the $\nu$ coordinate and an illustrative figure are provided in Section~\ref{sisec:nu} of the Supplementary Materials.
		We first study the influence of NQEs on the extent of proton transfer in the flat-rhombic phase at 4\,GPa.
		Within a classical description of nuclear motion, protons remain bound to their nearest oxygen atoms on the (nanosecond) timescale of our simulations.
		This remains true all the way up to the phase transition to the hexatic phase near 300 K.
		This can be seen clearly in the red curves in Fig.~\ref{fig:proton_disorder}(b), which never reach $\nu=0$.
		Since we do not observe any proton transfer events within the timescale of our simulations, we are unable to estimate the classical free energy barrier.
		On the other hand, our simulations that account for NQEs exhibit low, finite free energy barriers for proton transfer in the flat-rhombic phase.
		This is true even in our lowest temperature simulations performed at 100\,K.
		From these results alone, it is clear that NQEs enable proton transfer in the flat-rhombic phase.
		To further characterize the mechanism behind the onset of this proton transfer, we compute the joint free energy distribution of the proton transfer coordinate $\nu$ and the distance between the two oxygen atoms involved in the proton transfer $d_{\text{OO}}$.
		The resulting two-dimensional profiles in Fig.~\ref{fig:proton_disorder}(c) exhibit two major differences between the classical and quantum cases.
		First, the range of $d_{\text{OO}}$ distances extend to much smaller values with the inclusion of NQEs, meaning that the oxygen atoms themselves are closer together in the quantum simulations.
		Second, even for the same $d_{\text{OO}}$ values, proton transfer is dramatically enhanced by NQEs when compared to our classical simulations.
		The net result is that NQEs significantly reduce the barrier for proton transfer from classically forbidden to readily shared.
		We next study the mean squared displacement of protons to check if the reduced proton transfer barrier results in long-range proton dynamics. 
		As shown in Fig.~\ref{fig:proton_disorder}(d), NQEs do not induce diffusion of protons, implying that the flat-rhombic phase is dynamically disordered but not superionic.
		This can be intuitively understood by comparing the classical and quantum snapshots in Fig.~\ref{fig:proton_disorder}(a).
		In the classical snapshot, the hydrogen bond network is ordered, in the sense that protons involved in hydrogen bonds are always covalently bonded to the oxygen above them.
		In the quantum snapshot, we can see that some protons are covalently bonded to the oxygen above, while others are bonded to the oxygen below.
		In these simulations, we see that the protons are dynamically shuttling back and forth between the same two water molecules, which explains the lack of long-range proton diffusion.
		It is worth noting that the ice VII phase of bulk water exhibits a similar behaviour at room temperature, albeit at much higher pressures in the 40-50 GPa range~\cite{benoit_tunnelling_1998}.
		Classical ice VIII is proton-ordered, while NQEs induce a phase transition into ice VII with facile sharing of protons between hydrogen-bonded oxygen atoms. 
		Compared to bulk, NQEs induce proton sharing in nanoconfined flat-rhombic ice at much lower pressures --  even down to 1\,GPa (see Section~\ref{sisec:pressure_flat_rhombic} of the Supplementary Materials). 
		This heightened degree of proton sharing at much lower pressures than in bulk likely arises from the short hydrogen bonds in the flat-rhombic phase~\cite{ravindra_quasi-one-dimensional_2023}, which lead to an increased capacity for proton transfer.

		\begin{figure}
			\centering
			\includegraphics[width=3.375in]{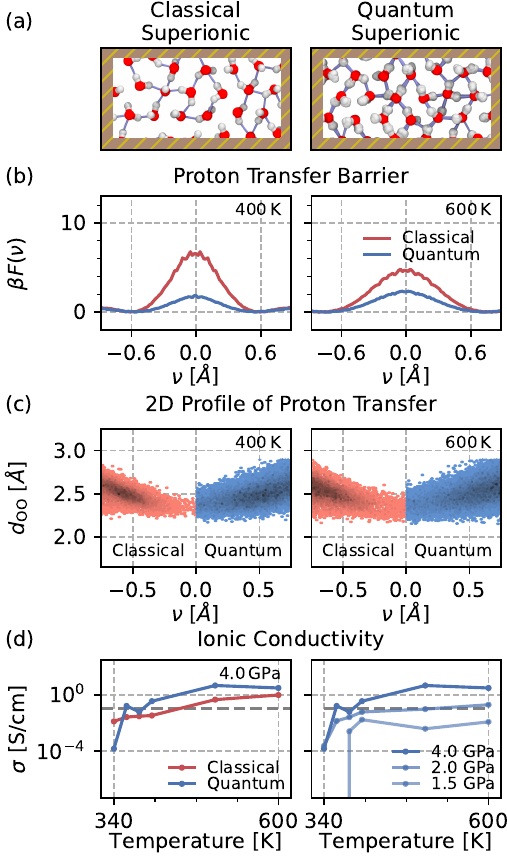}
			\caption{\textbf{Enhanced ionic conductivity in hexatic water.} (a) Snapshots of the superionic hexatic phase without (left) and with (right) nuclear quantum effects. The snapshot on the right shows all 32 beads from our path integral molecular dynamics simulations, overlaid on top of each other. No substantial qualitative difference is visible. (b) Free energy profiles along the proton transfer coordinate $\nu$ for the hexatic phase without (red) and with (blue) nuclear quantum effects at 100\,K (left) and 300\,K (right). (c) The joint free energy profile of neighbouring oxygen-oxygen distances $d_{\text{OO}}$ and the proton transfer coordinate $\nu$ for the hexatic phase. Darker regions correspond to lower free energies. By definition, the distributions are symmetric about $\nu$, so we only show the classical (red) distributions for negative $\nu$ values and the quantum (blue) distributions for positive $\nu$ values. (d) The left panel shows the classical and quantum ionic conductivities of the hexatic phase at 4\,GPa. The right panel shows the quantum ionic conductivity at 1.5, 2.0 and 4.0\,GPa. For both panels, the 0.1 S/cm threshold for superionic behaviour is shown with a dashed grey line. Error bars are omitted for clarity here but can be found in Fig.~\ref{sifig:superionic_cond} of the Supplementary Materials.}
			\label{fig:ionic_conductivity}
		\end{figure}
		
		
		%
		The differences caused by NQEs in the hexatic phase at 4\,GPa are perhaps even more marked.
		Ref.~\citenum{kapil_first-principles_2022} estimates the ionic conductivity of the hexatic phase using classical linear response theory.
		At around 400\,K, the ionic conductivity is shown to exceed 0.1 S/cm, which is the threshold for superionic behaviour that we will use in this work as well~\cite{kreuer_proton_1996,kapil_first-principles_2022}.
		Fig~\ref{fig:ionic_conductivity}(a) shows snapshots that compare our classical and quantum simulations at 400\,K.
		These snapshots do not exhibit any appreciable qualitative difference between classical and quantum superionic hexatic water.
		However, Fig.~\ref{fig:ionic_conductivity}(b) shows that NQEs lower the proton transfer barrier by a factor of four at 400\,K and by a factor of around two at 600\,K.
		Hence, NQEs do cause quantitative differences in proton transfer in the hexatic phase.
		We again compute the joint free energy distribution of the proton transfer coordinate $\nu$ and the oxygen-oxygen distance $d_{\text{OO}}$, as shown in Fig.~\ref{fig:ionic_conductivity}(c).
		The changes to the $d_{\text{OO}}$ distance distribution between the classical and quantum simulations are much more modest in the hexatic phase than the changes observed in the flat-rhombic phase.
		However, the inclusion of NQEs still leads to an enhancement in proton transfer for a given $d_{\text{OO}}$ value, just as we saw previously.
		Unlike in the flat-rhombic phase, this proton transfer is associated with long-range proton diffusion in the hexatic phase (see Section~\ref{sisec:pressure_hexatic} of the Supplementary Materials).
		This directly suggests an increase in the overall ionic conductivity. 
		To confirm this, we calculate the classical and quantum ionic conductivity in Fig.~\ref{fig:ionic_conductivity}(d) using the Einstein-Helfand relations with classical and centroid molecular dynamics, respectively.
		With the inclusion of NQEs, the ionic conductivity of the disordered phase increases by over a factor of 10, which is likely due to zero-point O--H bond fluctuations.
		This surge in proton dissociation reduces the onset temperature for superionic behaviour from approximately 400\,K to 360\,K.
		%
		
		
		%
		This NQE-induced enhancement of the ionic conductivity also occurs at lower lateral pressures.
		This is evidenced by the plots on the right side of Fig.~\ref{fig:ionic_conductivity}(d), which show the ionic conductivity of monolayer water across the entire hexatic and liquid regime in the 1-4\,GPa range.
		Estimating the classical and quantum ionic conductivity at low temperatures and pressures is statistically challenging due to the high variance associated with the microscopic ionic flux~\cite{grasselli_topological_2019}.
		This can result in erroneously large values of ionic conductivities when the O--H bonds are bound at low temperatures and pressures. 
		However, considering that superionic conductivity is driven by O--H dissociation, we estimate the ionic conductivity only for state points where we observe O--H dissociation events.
		The full details of our analysis are reported in Section~\ref{sisec:pressure_hexatic} of the Supplementary Materials
		The role of NQEs in these settings is stark.
		A large portion of the disordered region in the P--T phase diagram, which was previously considered bound, becomes either partially dissociated or superionic with the inclusion of NQEs.
		The superionic region within the phase diagram is greatly expanded, with nearly the entire disordered region above 2\,GPa exhibiting superionicity.
		Even for 1\,GPa conditions where we do not observe superionic behaviour, water becomes partially dissociated within the nanosecond timescale.
		%
		
		
		%
		\textit{Conclusions.} We have used machine learning-based first-principles simulations to investigate the influence of NQEs on proton transfer in nanoconfined water. 
		In the solid flat-rhombic phase, NQEs lead to a dynamic disordering of protons in which protons shuttle back and forth between neighbouring oxygen atoms.
		As such events are classically forbidden, this dynamical disorder may be attributed to the tunnelling of protons.
		Importantly, this shuttling behaviour does not lead to proton diffusion over long distances.
		In the hexatic phase, NQEs dramatically enhance the ionic conductivity of water, reducing the onset conditions of superionic behaviour from 4 to 2\,GPa and inducing partial O--H dissociation down to 1\,GPa.
		Considering that water encapsulated between graphene sheets experiences lateral pressures on the order of 1\,GPa, our simulations suggest that current experimental setups could be adapted to probe superionic or dissociated water.
		Indeed, recent work by ~\citet{wang_plane_2024} observes high ionic conductivity for water in nanocapillaries of width below 1 nanometer.
		The role of NQEs in inducing this enhanced ionic conductivity could be experimentally probed by studying the in-plane ionic conductivity of water isotopomers.
		Our work highlights the general importance of incorporating NQEs in simulations of nanofluidic transport, even in thermodynamic conditions where the effects of quantum nuclear motion are expected to be small in the corresponding bulk systems. 
		%
		
		\section*{Acknowledgements}
		P.R. thanks David R. Reichman for his academic support and the Winston Churchill Foundation of the United States for their financial support.
		X.R.A, S.W.C, and A.M. acknowledge support from the European Union under the ``n-AQUA" European Research Council project (Grant no. 101071937).
		BXS acknowledges support from the EPSRC Doctoral Training Partnership (EP/T517847/1).
		V.K acknowledges the Ernest Oppenheimer Early Career Fellowship and the Sydney Harvey Junior Research Fellowship, Churchill College, University of Cambridge, as well as startup funds from UCL.
		We are grateful for computational support from the Swiss National Supercomputing Centre under project s1209, the UK national high-performance computing service, ARCHER2, for which access was obtained via the UKCP consortium and the EPSRC grant ref EP/P022561/1, the Cambridge Service for Data Driven Discovery (CSD3) and the Cirrus UK National Tier-2 HPC Service at EPCC (\href{http://www.cirrus.ac.uk}{http://www.cirrus.ac.uk}) funded by the University of Edinburgh and EPSRC (EP/P020267/1)
		P.R. acknowledges that: ``This material is based upon work supported by the U.S. Department of Energy, Office of Science, Office of Advanced Scientific Computing Research, Department of Energy Computational Science Graduate Fellowship under Award Number DE-SC0024386.''
		P.R. also makes the disclaimer that: ``This report was prepared as an account of work sponsored by an agency of the United States Government. Neither the United States Government nor any agency thereof, nor any of their employees, makes any warranty, express or implied, or assumes any legal liability or responsibility for the accuracy, completeness, or usefulness of any information, apparatus, product, or process disclosed, or represents that its use would not infringe privately owned rights. Reference herein to any specific commercial product, process, or service by trade name, trademark, manufacturer, or otherwise does not necessarily constitute or imply its endorsement, recommendation, or favoring by the United States Government or any agency thereof. The views and opinions of authors expressed herein do not necessarily state or reflect those of the United States Government or any agency thereof.''
		

%


		\appendixformat
		\onecolumngrid 
		
		\newpage
		
		\section*{Supplementary Materials}
		
		This document provides additional details and analysis to support the results in the main text.
		Section~\ref{sisec:comp_details} provides a detailed description of our nanoconfined water simulation setup.
		In Section~\ref{sisec:benchmark}, we assess the accuracy of the revPBE0-D3 functional for describing O--H dissociation events.
		Section~\ref{sisec:nu} contains a rigorous definition of the proton transfer coordinate $\nu$ that we use to quantify proton transfer.
		Sections~\ref{sisec:pressure_flat_rhombic} and~\ref{sisec:pressure_hexatic} show the pressure dependence of proton transfer in the flat-rhombic and hexatic phases, respectively.
		
		\section{Computational details}
		\label{sisec:comp_details}
		
		The total potential energy for our simulations is the sum of a water-carbon confining potential and a water-water machine learning interatomic potential (MLIP).
		The confining walls of our system are uniform planes that are kept at a fixed separation width of 5\AA{}.
		The water molecules form a monolayer plane between these two walls.
		The interaction between the water molecules and the graphene sheets is treated as a Morse potential that acts uniformly in the direction perpendicular to the plane of water molecules.
		Hence, each atom in our system always experiences two Morse potentials -- one from each confining wall.
		The functional form of this potential $V_{morse}$ is:
		
		\begin{equation}
			V_{morse} = D_0 \qty[\qty(1 - e^{-a(r-r_e)})^2 - 1]
		\end{equation}
		
		\noindent
		In the above, $r$ is the distance of the atom from the wall.
		The values of the above parameters are taken from a previous study that fit a Morse potential to match water-carbon interaction energies computed at the quantum Monte Carlo level~\cite{chen_evidence_2016}.
		These values are $D_0 = 5.78 \cross 10^{-2} \text{ eV}$, $r_e = 3.85$~\AA{}, and $a = 0.92 $~\AA{}${}^{-1}$.
		For the water-water interactions, we use the same MLIP as in Ref.~\citenum{kapil_first-principles_2022}.
		The reported root-mean-square energy error of this MLIP is 2.4 meV per water molecule, and the reported force error is 75 meV/\AA{} per water molecule.
		In Section~\ref{sisec:benchmark}, we further characterize the accuracy of our MLIP for describing O--H dissociation events specifically. \\
		
		Our simulation cells contain 144 water molecules, which has previously been shown to be sufficiently large to avoid finite size effects for this system~\cite{kapil_first-principles_2022,ravindra_quasi-one-dimensional_2023}.
		Furthermore, due to the short-ranged nature of nuclear quantum effects (NQEs), we believe that this choice remains reasonable for the purposes of our analysis.
		We perform simulations at lateral pressures of 1.0, 1.5, 2.0, 3.0, and 4.0\,GPa.
		For each lateral pressure, we perform simulations at temperatures from 200\,K to 600\,K at 20\,K intervals.
		Classical simulations are performed with a timestep of 0.5 femtoseconds with an optimal sampling generalized Langevin equation thermostat. 
		These simulations are performed in the $NP_{\textrm{lat}}$T ensemble, with a modified Martyna-Tuckerman-Tobias-Klien barostat~\cite{kapil_first-principles_2022}.
		A constraint on the lattice vector perpendicular to the lateral plane is used to apply the lateral pressure. 
		Dynamical simulations use a time step of 0.5 femtoseconds and a stochastic velocity rescaling thermostat in the $NVT$ ensemble. 
		Positions are sampled every 100 femtoseconds to estimate transport quantities.
		For all temperatures and pressures, our path-integral simulations use 32 imaginary-time slices with a PILE thermostat~\cite{ceriotti_efficient_2012} and a timestep of 0.5\,femtoseconds.
		The path integral coarse graining simulations (PIGS)~\cite{musil_quantum_2022} were performed following the protocol illustrated in Ref.~\citenum{kapil_first-principles_2024}, wherein the quantum effective potential was estimated from a path integral molecular dynamics simulation at 500\,K, within the temperature elevation (Te) approximation.
		This effective potential was then used to estimate the vibrational spectra of monolayer hexatic water. 
		Since the Te \textit{ansatz} recovers the correct $\omega \to 0$ limit of the vibrational density of states, a single PIGS effective potential can be used to estimate transport quantities across a range of temperatures.
		Further details on the simulation setup can be found in the input files (to be provided when the manuscript is accepted). \\
		%
		
		In Fig.~\ref{fig:phase_diagram} of the main text, we provide the phase diagrams of nanoconfined water using both classical and quantum nuclei.
		To compute these phase diagrams, we estimate the temperature of phase boundaries at each simulated lateral pressure.
		The boundary between the flat-rhombic and hexatic phases is defined as the temperature at which the flat-rhombic phase ceases to be metastable on the timescale of our simulations, which is approximately 100 picoseconds.
		Hence, the displayed phase boundary provides an overestimate of the true phase boundary temperature.
		The hexatic-liquid phase transition has previously been shown to be continuous~\cite{kapil_first-principles_2022}, so we expect the phase transition to be observed nearly instantly, even on our limited simulation timescales.
		The boundary between the hexatic and superionic phases is chosen as the lowest temperature for which we observe ionic conductivities above 0.1 S/cm.
		This is the same criterion that we have used for superionic proton transport throughout this manuscript, and it is consistent with previous work~\cite{kreuer_proton_1996,kapil_first-principles_2022}.
		Finally, the errors in conductivities presented in Fig.~\ref{sisec:pressure_hexatic} are calculated using the method described in Ref.~\citenum{mccluskey2024accurateestimationdiffusioncoefficients} using the kinisi software~\cite{McCluskey_2024}.
		
		\section{Benchmarking revPBE0-D3 for bond breaking}
		\label{sisec:benchmark}

		\begin{figure}[t!]
			\centering
			\includegraphics[width=\textwidth]{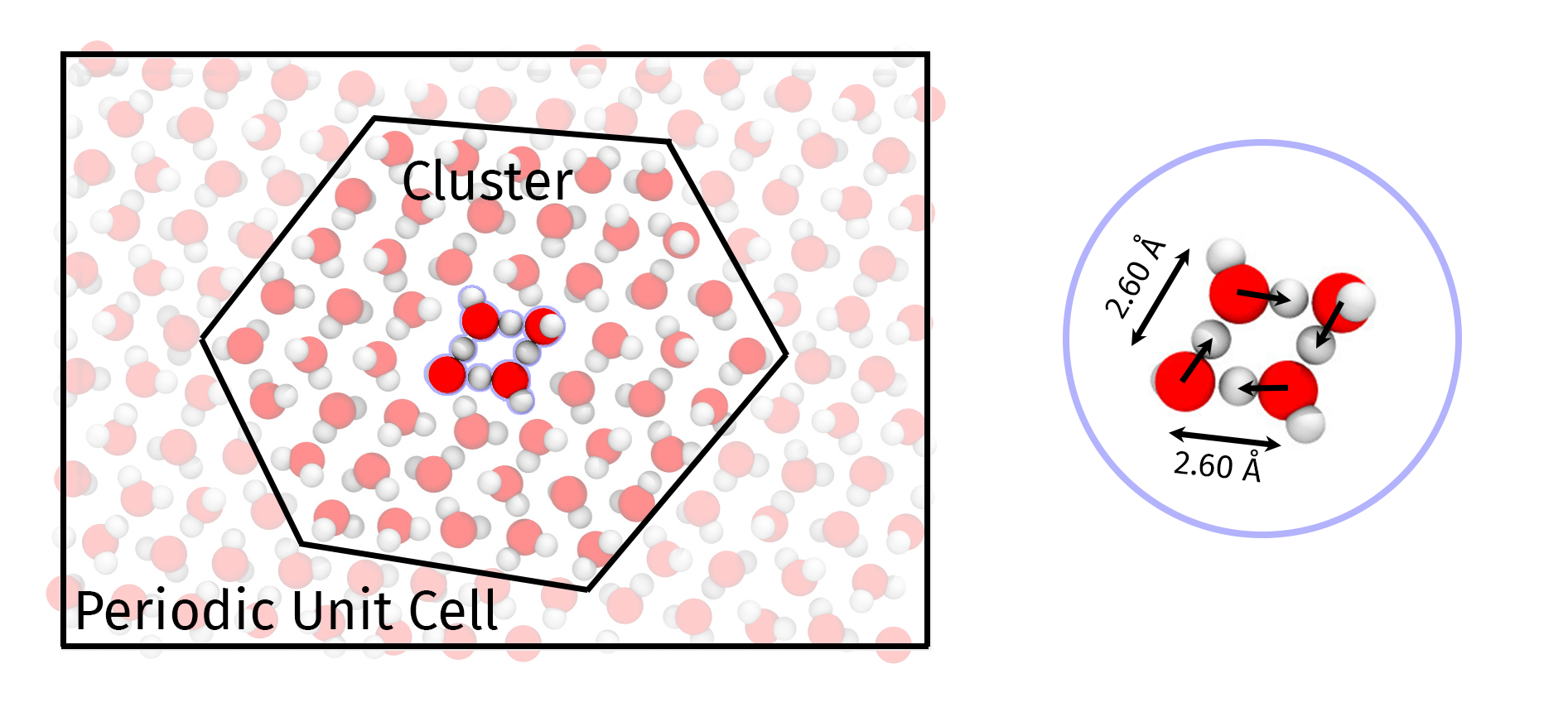}
			\caption{\textbf{The setup used to benchmark DFT functionals.} The left image shows the full periodic unit cell, as well as an example of a cluster of water molecules that are used for our ``cluster'' calculations. The four central water molecules that are highlighted in blue are shown again in the image on the right. The O--H bond lengths in these highlighted molecules are increased gradually in order to replicate proton transfer. The directions along which the O--H bond length is increased are indicated by the smaller arrows. The parallelogram formed by the four water molecules has an O--O distance of $2.60\,$\AA{} and a smaller vertex angle of $58.3^{\circ}$.}
			\label{sifig:water_tetramer}
		\end{figure}
		
		\begin{figure}[t!]
			\centering
			\includegraphics[width=3.375in]{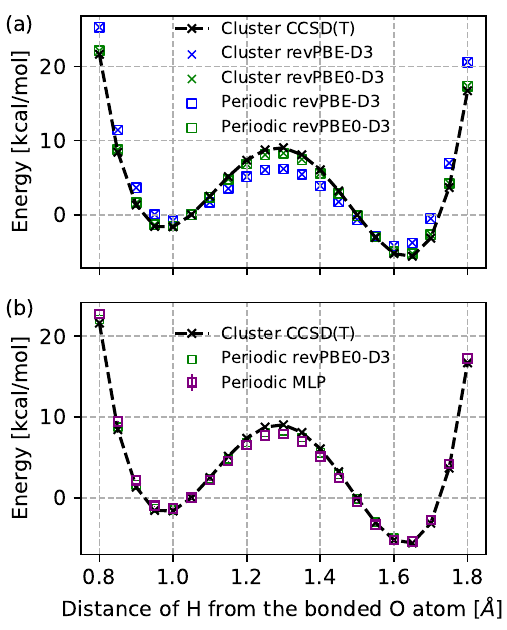}
			\caption{\textbf{Benchmarking density functionals and the MLIP against CCSD(T) for proton transfer.} (a) The energy along the proton hopping barrier in monolayer water estimated at different levels of theory, with both periodic unit cell and cluster calculations. (b) The trained MLIP's proton hopping barrier closely follows that of the underlying revPBE0-D3 functional. Both the MLIP and revPBE0-D3 functional match the forward barrier of CCSD(T) within chemical accuracy. All energies are relative to the bound state with an O--H distance of 1.1\,\AA{}.}
			\label{sifig:benchmark}
		\end{figure}

		To benchmark the performance of the revPBE0-D3 functional (with zero-damping) for bond breaking, we compare its performance to calculations at the level of coupled cluster theory with single, double, and perturbative triple excitations (CCSD(T)) with a local approximation.
		We start with a snapshot of the monolayer superionic phase from our classical simulations.
		An example periodic unit cell is shown in Fig.~\ref{sifig:water_tetramer}.
		We then select four adjacent water molecules, highlighted in blue, that form a parallelogram-shaped hydrogen bond network.
		For each selected water molecule, the hydrogen atoms that form the parallelogram are moved incrementally along their hydrogen bond directions in 0.1 \AA{} steps, as indicated by the small arrows in the circled image on the right of Fig.~\ref{sifig:water_tetramer}.
		At each increment, the positions of the four water molecules are relaxed using our MLIP, keeping all the other atoms fixed.
		Single-point calculations are performed along this trajectory at various levels of theory and are summarized in Fig.~\ref{sifig:benchmark}. 
		The density functional theory (DFT) and MLIP calculations are performed on the periodic unit cell, whereas CCSD(T) calculations are performed on a cluster that has been cleaved out from the unit cell, as shown in Fig.~\ref{sifig:water_tetramer}.
		In Fig.~\ref{sifig:benchmark}(a), we show that the cluster forward barrier closely matches the periodic unit cell forward barrier within 0.04\,kcal/mol. \\

		Overall, we find that the revPBE0-D3 functional agrees with CCSDT(T) on the forward barrier to around 0.5\,kcal/mol.
		On the other hand, the revPBE-D3 functional underestimates the barrier for proton transfer by around 4\,kcal/mol.
		Additionally, the MLIP from Ref.~\citenum{kapil_first-principles_2022} agrees with revPBE0-D3 to around 0.25\,kcal/mol, meaning that the MLIP faithfully reproduces the predictions of revPBE0-D3, even during O--H dissociation events.
		Hence, both the revPBE0-D3 functional and the MLIP agree with CCSD(T) within chemical accuracy. 
		As a final comment, it is worth noting that the snapshots are not geometry optimized at CCSD(T), which means that the reference barrier is an overestimate. \\

		DFT calculations are performed within CP2K~\cite{kuhneCP2KElectronicStructure2020} while the cluster CCSD(T) calculations are performed with MRCC~\cite{kallayMRCCProgramSystem2020} using the local natural orbital (LNO) approximation to CCSD(T).
		The CP2K calculations were performed using the Goedecker-Teter-Hutter (GTH) pseudopotentials~\cite{goedeckerSeparableDualspaceGaussian1996}, with the corresponding TZV2P-GTH basis sets.
		revPBE0-D3 was performed using the auxiliary-density-matrix method~\cite{guidonAuxiliaryDensityMatrix2010} for approximating the Fock exchange matrix and energy with the cpFIT3 auxiliary basis set. 
		We perform a two-point extrapolation~\cite{neeseRevisitingAtomicNatural2011} of the aug-cc-pVTZ and aug-cc-pVQZ Dunning family basis sets~\cite{petersonAccurateCorrelationConsistent2002} with `normal' LNO thresholds and make a subsequent correction to the `tight' thresholds performed with the aug-cc-pVTZ basis set.
		
		\clearpage
		
		\section{The Proton Transfer Coordinate}
		\label{sisec:nu}
		
		\begin{figure*}[h!]
			\centering
			\includegraphics[width=3.375in]{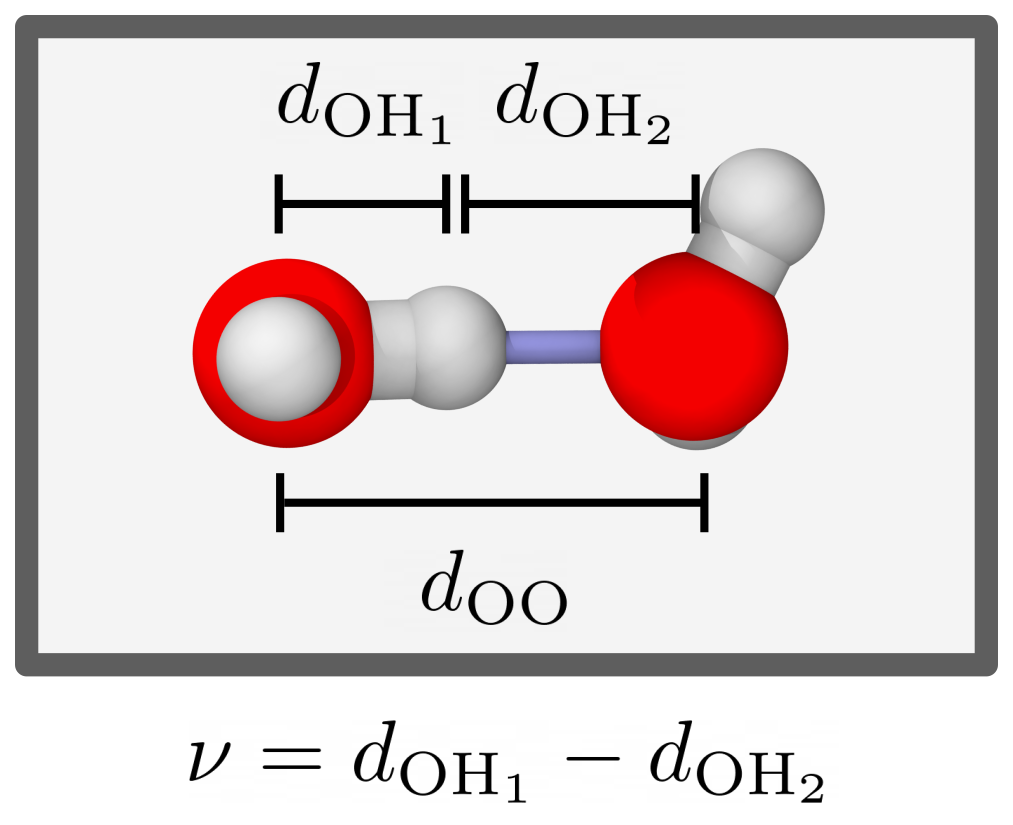}
			\caption{\textbf{Illustration of the proton transfer coordinate $\nu$.} For a given proton (in white), $d_{\text{OH}_1}$ and $d_{\text{OH}_2}$ are the scalar distances to the two nearest oxygen atoms (in red). The choice for which oxygen atom corresponds to $d_{\text{OH}_1}$ or $d_{\text{OH}_2}$ is arbitrary, meaning that the sign of $\nu$ is also arbitrary.}
			\label{sifig:proton_transfer_coordinate}
		\end{figure*}
		
		In the main text, we use the proton transfer coordinate $\nu$ to characterize proton transfer between water molecules. Fig.~\ref{sifig:proton_transfer_coordinate} shows how this coordinate is defined and computed. For each proton, we find the two nearest oxygen atoms. We denote the distance from the proton to each of the oxygen atoms as $d_{\text{OH}_1}$ and $d_{\text{OH}_2}$. The proton transfer coordinate $\nu$ is then given by the difference of these two scalar distances. Since the assignment of $d_{\text{OH}_1}$ and $d_{\text{OH}_2}$ is arbitrary, the sign of $\nu$ is arbitrary, and only the absolute value of $\nu$ should be assigned any meaning. \\
		
		Under this definition, values near $\nu=0$ correspond to protons that are equidistant from their two neighbouring oxygen atoms, whereas values of $\nu$ with large absolute values correspond to protons that are much closer to their nearest oxygen atom than to their second nearest oxygen atom. Importantly, for a proton to transfer from one oxygen to another, it must pass through $\nu=0$, which allows us to use the free energy barrier along the proton transfer coordinate to understand the kinetics of proton transfer.
		
		\clearpage
		
		\section{Pressure-dependent proton disorder in flat-rhombic ice}
		\label{sisec:pressure_flat_rhombic}
		
		\begin{figure}[h!]
			\centering
			\includegraphics[width=3.375in]{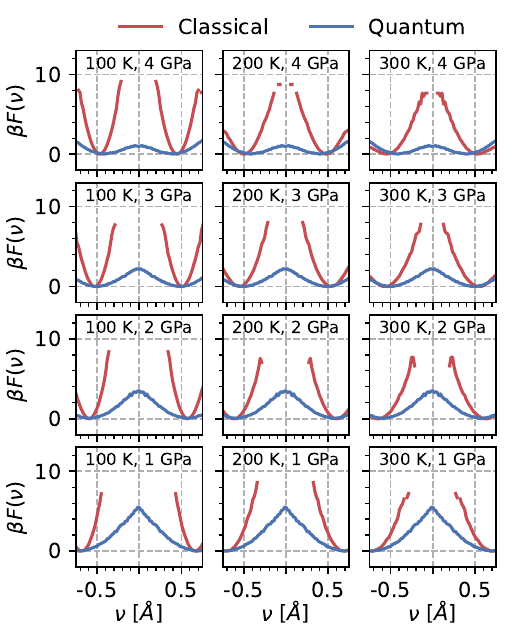}
			\caption{\textbf{The temperature scaled free energy along the proton transfer coordinate for the flat-rhombic phase.} Each panel corresponds to a different thermodynamic state point. The classical and quantum free energy profiles are shown in red and blue, respectively.}
			\label{sifig:rhombic_ptc}
		\end{figure}

		Fig.~\ref{sifig:rhombic_ptc} reports the classical and quantum proton transfer free energy profiles for the flat rhombic phase from 1-4\,GPa and 100-300\,K. 
		On the timescales of our simulations, we do not observe proton transfer in any of our classical simulations of the flat-rhombic phase.
		Accordingly, we do not estimate the height of the classical free energy barrier.
		On the other hand, when NQEs are accounted for, we observe proton transfer at all of the selected state points, even down to 100\,K and 1\,GPa.
		These temperature and lateral pressure conditions are certainly accessible to graphene nanocapillary experiments~\cite{algara-siller_square_2015}.
		As discussed in the main text, these proton transfer events in the flat-rhombic phase involve protons shuttling back and forth between adjacent water molecules, without any diffusive proton transport.
		
		\clearpage

		\section{Proton diffusion in hexatic water}
		\label{sisec:pressure_hexatic}
		
		\vspace*{-1em}
		
		\begin{figure}[h!]
			\centering
			\includegraphics[width=3.375in]{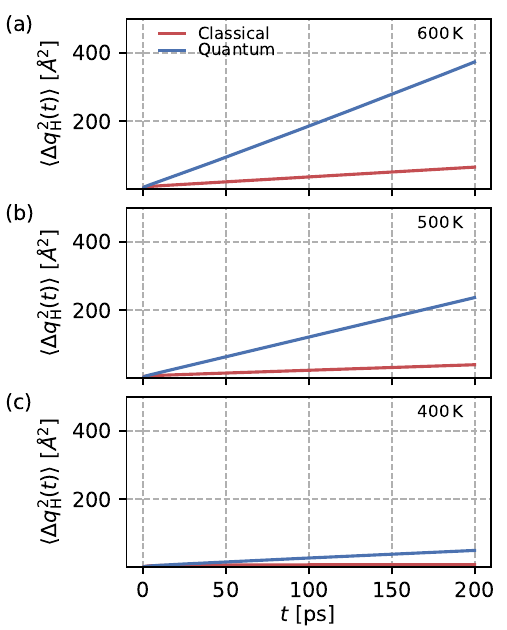}
			\caption{\textbf{Mean squared displacement of protons in the hexatic phase at 4\,GPa.} The classical and quantum mean squared displacements are shown in red and blue, respectively.}
			\label{sifig:proton_diff_hexatic}
		\end{figure}

		Fig.~\ref{sifig:proton_diff_hexatic} reports the classical and quantum proton mean squared displacements for the hexatic phase at 4\,GPa and 400-600\,K. 
		Over the same amount of time, NQEs lead to over an order of magnitude increase in the mean squared displacement for protons. \\
		
\begin{figure*}
			\centering
			\includegraphics[width=3.375in]{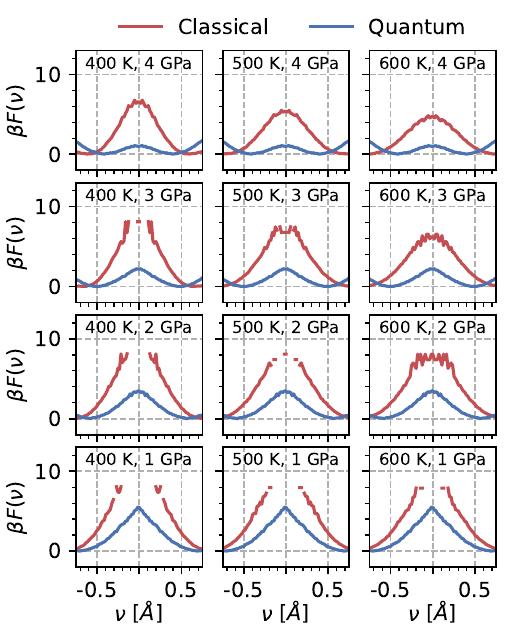}
			\caption{\textbf{Analyses for identifying conditions which exhibit O--H dissociation.} The classical and quantum free energy profiles along the proton transfer coordinate $\nu$ are shown in red and blue, respectively. Each panel corresponds to a different thermodynamic state point.}
			\label{sifig:superionic_barriers}
		\end{figure*}

		The Einstein-Helfand relation for ionic conductivity is given by $ \sigma = \frac{\beta}{3 V}\lim_{t \to \infty} \frac{\langle\left|\Delta \mu(t)\right|^2\rangle}{2 t}$, where $V$ is the volume and $\mu$ is the total dipole moment, with partial charges assigned equal to the oxidation number of the atoms. 
		Due to the $\beta$ prefactor, any fixed statistical error in the mean squared displacement of the total dipole moment leads to a linearly increasing statistical error in the ionic conductivity. 
		At low temperatures, this relation suffers from poor convergence, requiring very long trajectories. 
		The $\beta$ prefactor can result in an erroneous increase in ionic conductivity with decreasing temperature when the simulations are not long enough.
		This makes it difficult to compute the temperature dependence of ionic conductivity accurately, especially when O--H bonds remain intact.
		This can result in an erroneous increase in ionic conductivity with decreasing temperature. To avoid these statistical issues in identifying superionic behaviour, we focus on the fact that ionic conduction in superionic water is driven by O--H bond dissociation. 
		A phase cannot be superionic without exhibiting dissociated O--H bonds.
		Hence, we plot the proton transfer barriers in Fig.~\ref{sifig:superionic_barriers} to check which simulations exhibit O--H dissociation.
		We estimate the ionic conductivity only for state points where O--H dissociation is observed.
		For these conditions, we achieve good statistical accuracy in estimating the ionic conductivity and can determine if the value exceeds the threshold of 0.1\,S/cm. \\
		
\begin{figure*}
			\centering
			\includegraphics[width=3.375in]{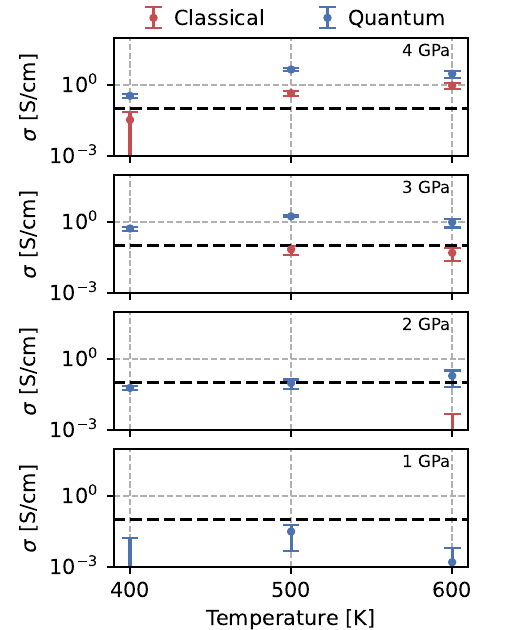}
			\caption{\textbf{The ionic conductivity of the hexatic phase across a variety of conditions.} The classical and quantum ionic conductivities at 1-4\,GPa for the temperatures that exhibit O--H dissociation on the timescale of our simulations. The horizontal dashed black line corresponds to the 0.1 S/cm cutoff that we use for superionic behaviour. For the data points where the error bars go off the plot, the error values are larger than the actual data values, meaning that the ionic conductivities are too close to 0 to provide a reliable estimate. Hence, the corresponding conditions clearly do not exhibit superionicity.}
			\label{sifig:superionic_cond}
		\end{figure*}

		Fig.~\ref{sifig:superionic_barriers} shows that our quantum simulations exhibit O--H dissociation at all state points in the 1-4\,GPa and 400-600\,K range.
		However, the classical simulations only exhibit O--H dissociation at the higher temperatures and lateral pressures.
		The ionic conductivities in Fig.~\ref{sifig:superionic_cond} show a similar general trend.
		Our classical simulations only exceed the 0.1\,S/cm threshold at 4\,GPa and 500-600\,K.
		However, the quantum simulations exhibit near-superionic proton transport in all simulations with a lateral pressure that is at least 2\,GPa.
		Even the 1\,GPa simulations show a non-trivial level of O--H dissociation, although they fail to meet the 0.1\,S/cm cutoff for superionicity.
		
	\end{document}